\documentclass[AMA,STIX1COL]{WileyNJD-v2}
\usepackage{booktabs}
\usepackage{multirow,hyperref}
\usepackage{footnote}
\usepackage{float}
\usepackage{comment}
\usepackage{url}

\articletype{Research Article}%

\received{26 April 2016}
\revised{6 June 2016}
\accepted{6 June 2016}

\newcommand{\OW}{\mbox{\tiny{OW}}}
\newcommand{\IPW}{\mbox{\tiny{IPW}}}
\newcommand{\AIPW}{\mbox{\tiny{AIPW}}}
\newcommand{\LR}{\mbox{\tiny{LR}}}
\newcommand{\DIF}{\mbox{\tiny{UNADJ}}}
\newcommand{\RR}{\mbox{\tiny{RR}}}
\newcommand{\RD}{\mbox{\tiny{RD}}}
\newcommand{\COR}{\mbox{\tiny{OR}}}

\newcommand{\sumi}{\sum_{i=1}^N}

\begin{document}

\title{Propensity score weighting for covariate adjustment in randomized clinical trials}

\author{Shuxi Zeng$^1$, Fan Li$^{2,*}$, Rui Wang$^{3,4}$, Fan Li$^1$}

\authormark{}

\address[1]{\orgdiv{Department of Statistical Science }, \orgname{Duke University}, \orgaddress{\state{North Carolina}, \country{USA}}}
\address[2]{\orgdiv{Department of Biostatistics}, \orgname{Yale School of Public Health}, \orgaddress{\state{Connecticut}, \country{USA}}}
\address[3]{\orgdiv{Department of Population Medicine}, \orgname{Harvard Pilgrim Health Care Institute and Harvard Medical School}, \orgaddress{\state{Massachusetts}, \country{USA}}} 
\address[3]{\orgdiv{Department of Biostatistics}, \orgname{Harvard T. H. Chan School of Public Health}, \orgaddress{\state{Massachusetts}, \country{USA}}} 

\corres{$^*$Fan Li\\
Department of Biostatistics\\
Yale School of Public Health\\
\email{fan.f.li@yale.edu}}

%\presentaddress{}

\abstract[Summary]{Chance imbalance in baseline characteristics is common in randomized clinical trials. Regression adjustment such as the analysis of covariance (ANCOVA) is often used to account for imbalance and increase precision of the treatment effect estimate. An objective alternative is through inverse probability weighting (IPW) of the propensity scores. Although IPW and ANCOVA are asymptotically equivalent, the former may demonstrate inferior performance in finite samples. In this article, we point out that IPW is a special case of the general class of balancing weights, and advocate to use overlap weighting (OW) for covariate adjustment. The OW method has a unique advantage of completely removing chance imbalance when the propensity score is estimated by logistic regression. We show that the OW estimator attains the same semiparametric variance lower bound as the most efficient ANCOVA estimator and the IPW estimator for a continuous outcome, and derive closed-form variance estimators for OW when estimating additive and ratio estimands. Through extensive simulations, we demonstrate OW consistently outperforms IPW in finite samples and improves the efficiency over ANCOVA and augmented IPW when the degree of treatment effect heterogeneity is moderate or when the outcome model is incorrectly specified. We apply the proposed OW estimator to the Best Apnea Interventions for Research (BestAIR) randomized trial to evaluate the effect of continuous positive airway pressure on patient health outcomes. All the discussed propensity score weighting methods are implemented in the R package \textbf{PSweight}.}


\jnlcitation{\cname{%
  \author{Zeng S., Li F., Wang R., Li F.}} (\cyear{2020}), 
  \ctitle{Propensity score weighting for covariate adjustment in randomized clinical trials}, \cjournal{Statistics in Medicine}, \cvol{0000}.}
\maketitle

\section{Introduction}

Randomized controlled trials are the gold standard for evaluating the efficacy and safety of new treatments and interventions. Statistically, randomization ensures the optimal internal validity and balances both measured and unmeasured confounders in expectation. This makes the simple unadjusted difference-in-means estimator unbiased for the intervention effect.\cite{Rosenberger2002} Frequently, important patient characteristics are collected at baseline; although over repeated experiments, they will be balanced between treatment arms, chance imbalance often arises in a single trial due to the random nature in allocating the treatment,\cite{Senn1989,Ciolino2015} especially when the sample size is limited.\cite{Thompson2015} If any of the baseline covariates are prognostic risk factors that are predictive of the outcome, adjusting for the imbalance of these factors in the analysis can improve the statistical power and provide a greater chance of identifying the treatment signals when they actually exist.\cite{Ciolino2015,Pocock2002,Hernandez2004}

%\comment{Existing methods for Covariate adjustment: regression and IPTW}

There are two general streams of methods for covariate adjustment in randomized trials: (outcome) regression adjustment\cite{yang2001efficiency,Kahan2016,leon2003semiparametric,tsiatis2008covariate,Zhang2008} and the inverse probability of treatment weighting (IPW or IPTW) based on propensity scores. \cite{williamson2014variance,shen2014inverse,Colantuoni2015} For regression adjustment with continuous outcomes, the analysis of covariance (ANCOVA) model is often used, where the outcome is regressed on the treatment, covariates and possibly their interactions.\cite{tsiatis2008covariate} The treatment effect is estimated by the coefficient of the treatment variable. With binary outcomes, a generalized linear model can be postulated to estimate the adjusted risk ratio or odds ratio, with the caveat that the regression coefficient of treatment may not represent the marginal effect due to non-collapsability.\cite{williamson2014variance}
Tsiatis and co-authors developed a suite of semiparametric ANCOVA estimators that improves efficiency over the unadjusted analysis in randomized trials.\cite{yang2001efficiency,leon2003semiparametric,tsiatis2008covariate} Lin \cite{lin2013agnostic} clarified that it is critical to incorporate covariate-by-treatment interaction terms in regression adjustment for efficiency gain. %The finite-sample performance of these ANCOVA estimators has been examined by previous simulations \cite{tsiatis2008covariate}. 
When the randomization probability is $1/2$, ANCOVA returns consistent point and interval estimates even if the outcome model is misspecified.\cite{yang2001efficiency,lin2013agnostic,wang2019analysis} However, misspecification of the outcome model can decrease precision in unbalanced experiments with treatment effect heterogeneity.\cite{Freedman2008} Another limitation of regression adjustment is the potential for inviting a `fishing expedition': one may search for an outcome model that gives the most dramatic treatment effect estimate which jeopardizes the objectivity of causal inference with randomized trials. \cite{tsiatis2008covariate,shen2014inverse}

Originally developed in the context of survey sampling and observational studies, \cite{lunceford2004stratification} IPW has been advocated as an objective alternative to ANCOVA in randomized trials.\cite{williamson2014variance} To implement IPW, one first fits a logistic \emph{working} model to estimate the propensity scores -- the conditional probability of receiving the treatment given the baseline covariates,\cite{Rosenbaum83} and then estimates the treatment effect by the difference of the weighted outcome -- weighted by the inverse of the estimated propensity -- between the treatment arms. In randomized trials, the treatment group is randomly assigned and the true propensity score is known. Therefore, the working propensity score model is always correctly specified, and the IPW estimator is consistent to the marginal treatment effect. For a continuous outcome, the IPW estimator with a logistic propensity model has the same large-sample variance as the efficient ANCOVA estimator, \cite{shen2014inverse,williamson2014variance} but it offers the following advantages. 

First, IPW separates the design and analysis in the sense that the propensity score model only involves baseline covariates and the treatment indicator; it does not require the access to the outcome and hence avoids the `fishing expedition.' As such, IPW offers better transparency and objectivity in pre-specifying the analytical adjustment before outcomes are observed. Second, IPW preserves the marginal treatment effect estimand with non-continuous outcomes, while the interpretation of the outcome regression coefficient may change according to different covariate specifications.\cite{Hauck1998,Robinson1991a} Third, IPW can easily obtain treatment effect estimates for rare binary or categorical outcomes whereas outcome models often fail to converge in such situations. \cite{williamson2014variance} This is particularly the case when the target parameter is a risk ratio, where log-binomial models are known to have unsatisfying convergence properties.\cite{Zou2004} On the other hand, a major limitation of IPW is that it may be inefficient compared to ANCOVA with limited sample sizes and unbalanced treatment allocations.\cite{Raad2020} 

In this paper, we point out that IPW is a special case of the general class of propensity score weights, called the balancing weights,\cite{li2018balancing} many members of which could be used for covariate adjustment in randomized trials. Within this class, we advocate to use the overlap weighting (OW). \cite{li2018balancing,li2019addressing,schneider2001,crump2006, li2019propensity} %for efficiency gain.%which has been shown previously, in the context of observational studies, to offer theoretical and empirical gain in variance estimation compared to IPW.
In the context of randomized trials, a particularly attractive feature of OW is that, if the propensity score is estimated from a logistic working model, then OW leads to \emph{exact mean balance} of any baseline covariate in that model, and consequently remove the chance imbalance of that covariate. As a propensity score method, OW retains the aforementioned advantages of IPW while offers better finite-sample properties (Section \ref{sec:OW}). In Section \ref{sec:theory}, we demonstrate that the OW estimator, similar as IPW, achieves the same semiparametric variance lower bound and hence is asymptotically equivalent to the efficient ANCOVA estimator for continuous outcomes. For binary outcomes, we further provide closed-form variance estimators of the OW estimator for estimating marginal risk difference, risk ratio and odds ratio, which incorporates the uncertainty in estimating the propensity scores and achieves close to nominal coverage in finite samples. Through extensive simulations in Section \ref{sec:simulations}, we demonstrate the efficiency advantage of OW under small to moderate sample sizes, and also validate the proposed variance estimator for OW. Finally, in Section \ref{sec:application} we apply the proposed method to the Best Apnea Interventions for Research (BestAIR) randomized trial and evaluate the treatment effect of continuous positive airway pressure (CPAP) on several clinical outcomes. 

%The proposed methods as well as the IPW and augmented-IPW estimators are implemented in the R package \textbf{PSweight} for general propensity score weighting analysis \cite{zhou2020psweight}, available on CRAN at \url{https://CRAN.R-project.org/package=PSweight}. 

\section{Propensity score weighting for covariate adjustment} \label{sec:OW}
\subsection{The Balancing Weights}
We consider a randomized trial with two arms and $N$ patients, where $N_1$ and $N_0$ patients are randomized into the treatment and control arm, respectively. Let $Z_{i}=z$ be the binary treatment indicator, with $z=1$ indicates treatment and $z=0$ control. Under the potential outcome framework,\cite{Neyman1923} each unit has a pair of potential outcomes $\{Y_{i}(1),Y_{i}(0)\}$, mapped to the treatment and control condition, respectively, of which only the one corresponding to the actual treatment assigned is observed. We denote the observed outcome as $Y_i=Z_iY_i(1)+(1-Z_i)Y_i(0)$. In randomized trials, a collection of $p$ baseline variables could be recorded for each patient, denoted by $X_i=( X_{i1},\ldots,X_{ip})^T$. 
Denote $\mu_z=E\{Y_i(z)\}$ and $\mu_z(x)=E\{Y_i(z)|X_i=x\}$
as the marginal and conditional expectation of the outcome in arm $z$ ($z=0,1$), respectively. A common estimand on the additive scale is the average treatment effect (ATE):
\begin{equation}
\label{def:ATE}
\tau=E\{Y_{i}(1)-Y_{i}(0)\}=\mu_1-\mu_0.
\end{equation}
We assume that the treatment $Z$ is randomly assigned to patients, where $\Pr(Z_i=1|X_i,Y_i(1),Y_i(0))=\Pr(Z_i=1)=r$, and $0<r<1$ is the randomization probability (see Web Appendix A for additional discussions on randomization). The most typical study design uses balanced assignment with $r={1}/{2}$. Other values of $r$ may be possible, for example, when there is a perceived benefit of the treatment, and a larger proportion of patients are randomized to the intervention. Under randomization of treatment and the consistency assumption, we have $\tau=E(Y_i|Z_i=1)-E(Y_i|Z_i=0)$, and thus the unadjusted difference-in-means estimator is:
\begin{equation}
\label{eq:unadjusted_est}
\hat{\tau}^{\DIF}=\frac{\sum_{i=1}^{N}Z_{i}Y_{i}}{\sumi Z_i}-\frac{\sum_{i=1}^{N}(1-Z_{i})Y_{i}}{\sumi (1-Z_i)}.   
\end{equation}

Below we generalize the ATE to a class of weighted average treatment effect (WATE) estimands to construct alternative weighting methods. Assume the study sample is drawn from a probability density $f(x)$, and let $g(x)$ denote the covariate distribution density of a \emph{target population}, possibly different from the one represented by the observed sample. The ratio $h(x)=g(x)/f(x)$ is called a \emph{tilting function},\citep{li2019propensity} which re-weights the distribution of the baseline characteristics of the study sample to represent the target population. We can represent the ATE on the target population $g$ by a WATE estimand:
\begin{equation}
\tau^h=E_g[Y_i(1)-Y_i(0)]=\frac{E\{h(x)(\mu_1(x)-\mu_0(x))\}}{E\{h(x)\}}.    
\end{equation}
In practice, we usually pre-specify $h(x)$ instead of $g(x)$. Most commonly $h(x)$ is specified as a function of the propensity score or simply a constant. The propensity score \citep{Rosenbaum83} is the conditional probability of treatment given the covariates, $e(x)=\Pr(Z_i=1|X_i=x)$. Under the randomization assumption, $e(x)=\Pr(Z_i=1)=r$ for any baseline covariate value $x$, and therefore as long as $h(x)$ is a function of the propensity score $e(x)$, different $h$ corresponds to the same target population $g$, and the WATE reduces to ATE, i.e. $\tau^h=\tau$. This is \emph{distinct from observational studies}, where the propensity scores are usually unknown and vary between units, and consequently different $h(x)$ corresponds to different target populations and estimands. \cite{thomas2020using} This special feature under randomized trials provides the basis for considering alternative weighting strategies to achieve better finite-sample performances. 

%Hirano et al. \cite{Hirano03} introduced a general class of weighted average treatment effect (WATE) estimands, which, as will be seen in due course, lead to a unified construction of different propensity score weighting methods. Specifically, 

In the context of confounding adjustment in observational studies, Li et al. \cite{li2018balancing} proposed a class of propensity score weights, named the \emph{balancing weights}, to estimate WATE. Specifically, given any $h(x)$, the balancing weights for patients in the treatment and control arm are defined as:
\begin{equation} \label{def:bw}
   w_1(x) =h(x)/{e(x)}, \quad w_0(x) =h(x)/\{1-e(x)\},
\end{equation}
which balances the distributions of the covariates between treatment and control arms in the target population, so that $f_1(x)w_1(x) = f_0(x)w_0(x) = f(x)h(x)$, where $f_z(x)$ is the conditional distribution of covariates in treatment arm $z$. \cite{wallace2015doubly,li2018balancing} Then, one can use the following H\'{a}jek-type estimator %\cite{Hajek1971} 
to estimate $\tau^h$:
\begin{equation}
\label{eq:sampleWATE}
\hat{\tau}^h=\hat{\mu}^h_1-\hat{\mu}^h_0=\frac{\sumi w_1(x_i)Z_i Y_i}{\sumi w_1(x_i)Z_i} -
              \frac{\sumi w_0(x_i)(1-Z_i) Y_i}{\sumi w_0(x_i)(1-Z_i)}.
\end{equation}
The function $h(x)$ can take any form, each corresponding to a specific weighting scheme. For example, when $h(x)=1$, the balancing weights become the inverse probability weights, $(w_1, w_0)=(1/e(x), 1/\{1-e(x)\})$; when $h(x)=e(x)(1-e(x))$, we have the overlap weights,\cite{li2018balancing} $(w_1, w_0)=(1-e(x), e(x))$, which was also independently developed by Wallace and Moodie  \cite{wallace2015doubly} in the context of dynamic treatment regimes. Other examples of the balancing weights include the average treatment effect among treated (ATT) weights \cite{hirano2001estimation} and the matching weights.\cite{li2013weighting} 

IPW is the most well-known case of the balancing weights. Specific to covariate adjustment in randomized trials, Williamson et al. \cite{williamson2014variance} and Shen et al. \cite{shen2014inverse} suggested the following IPW estimator of $\tau$:
\begin{equation}
\label{eq:IPW}
\hat{\tau}^{\IPW}=\frac{\sumi Z_i Y_i/\hat{e}_i}{\sumi Z_i/\hat{e}_i} -
              \frac{\sumi (1-Z_i) Y_i/(1-\hat{e}_i)}{\sumi (1-Z_i)/(1-\hat{e}_i)}.
\end{equation}
We will point out in Section \ref{sec:theory} that their findings on IPW are generally applicable to the balancing weights as long as $h(x)$ is a smooth function of the true propensity score. The choice of $h(x)$, however, will affect the finite-sample operating characteristics of the weighting estimator. In particular, below we will closely examine  the overlap weights. %due to its perceived efficiency advantages. % as evidenced by Li et al. \cite{li2019addressing} in a different context of observational data.

\subsection{The Overlap Weights\label{sec:OW_intro}}
% The overlap weights, $(w_1, w_0)=(1-e(x), e(x))$, weigh each patient by its probability of being assigned to the opposite group, and thus gradually down-weigh the patients whose propensity score are away from the center (0.5). 
%\shuxi{The overlap weights, $(w_1, w_0)=(1-e(x), e(x))$, weigh each patient by its probability of being assigned to the opposite group, and thus gradually down-weigh the patients whose probability of being assigned to the opposite treatment is lower.}
% In contrast to IPW, the patients with extreme (close to 0 and 1) propensity scores contribute the least in the estimation of the treatment effect. 

In observational studies, the overlap weights correspond to a target population with the most overlap in the baseline characteristics, and have been shown theoretically to give the smallest asymptotic variance of $\hat{\tau}^h$ among all balancing weights \cite{li2018balancing} as well as empirically reduce the variance of $\tau^{h}$ in finite samples.\cite{li2019addressing} Illustrative examples of the overlap population distribution can be found in Figure 1 of Li et al. \cite{li2018balancing} with a single covariate as well as in the bubble plot of Thomas et al. \cite{thomas2020overlap} with two covariates. In randomized trials, as discussed before, because the true propensity score is constant, the overlap weights and IPW target the same population estimand $\tau$, but their finite-sample operating characteristics can be markedly different, as elucidated below.   %For randomized trials, the true tilting function in each of the above examples is a constant because the true propensity score is a constant. 

The OW estimator for the ATE in randomized trials is
\begin{equation}
\label{eq:OWATE}
\hat{\tau}^{\OW}=\hat{\mu}_1-\hat{\mu}_0=\frac{\sumi (1-\hat{e}_i)Z_i Y_i}{\sumi (1-\hat{e}_i)Z_i} -
              \frac{\sumi \hat{e}_i(1-Z_i) Y_i}{\sumi \hat{e}_i(1-Z_i)},
\end{equation} 
where $\hat{e}_i=e(X_i;\hat{\theta})$ is the estimated propensity score from a working logistic regression model: 
\begin{equation}\label{eq:ps_logistic}
e_i={e}(X_i;\theta)=\frac{\exp(\theta_0+X_i^T\theta_1)}{1+\exp(\theta_0+X_i^T\theta_1)},  
\end{equation}
with parameters $\theta=(\theta_0,\theta_1^T)^T$ and $\hat{\theta}$ is the maximum likelihood estimate of $\theta$. Regarding the selection of covariates in the propensity score model, the previous literature suggests to include stratification variables as well as a small number of key prognostic factors pre-specified in the design stage.\cite{Raab2000, williamson2014variance} These guidelines are also applicable to the OW estimator. 

The logistic propensity score model fit underpins a unique \emph{exact balance} property of OW. Specifically, the overlap weights estimated from model \eqref{eq:ps_logistic} lead to exact mean balance of any predictor included in the model (Theorem 3 in Li et al. \cite{li2018balancing}):
\begin{equation}
\label{eq:balance}
\frac{\sumi (1-\hat{e}_i)Z_i X_{ji}}{\sumi (1-\hat{e}_i)Z_i} -
              \frac{\sumi \hat{e}_i(1-Z_i) X_{ji}}{\sumi \hat{e}_i(1-Z_i)}=0, \quad \mbox{for } j=1,...,p. 
\end{equation}
This property has important practical implications in randomized trials, namely, for any baseline covariate included in the propensity score model, the associated chance imbalance in a single randomized trial vanishes once the overlap weights are applied. If one reports the weighted mean differences in baseline covariates between arms (frequently included in the standard ``Table 1" in primary trial reports), those differences are identically zero. Thus the application of OW enhances the face validity of the randomized study.

More importantly, the exact mean balance property translates into better efficiency in estimating $\tau$. To illustrate the intuition, consider the following simple example. Suppose the true outcome surface is $Y_i=\alpha+Z_i\tau+X_i^T\beta_0+\varepsilon_i$ with $E(\varepsilon_i|Z_i,X_i)=0$. Denote the weighted chance imbalance in the baseline covariates by
\begin{equation*}
\Delta_X(w_0,w_1)=\frac{\sumi w_1(X_i)Z_i X_i}{\sumi w_1(X_i)Z_i} -
              \frac{\sumi w_0(X_i)(1-Z_i) X_i}{\sumi w_0(X_i)(1-Z_i)},
\end{equation*}
and the weighted difference in random noise by
\begin{equation*}
\Delta_\varepsilon(w_0,w_1)=\frac{\sumi w_1(X_i)Z_i \varepsilon_i}{\sumi w_1(X_i)Z_i} -
              \frac{\sumi w_0(X_i)(1-Z_i) \varepsilon_i}{\sumi w_0(X_i)(1-Z_i)}.
\end{equation*}
For the unadjusted estimator, substituting the true outcome surface in equation \eqref{eq:unadjusted_est} gives $\hat{\tau}^{\DIF}-\tau=\Delta_X(1,1)^T\beta_0+\Delta_\varepsilon(1,1)$.% where $\Delta_X(1,1)$ measures the chance imbalance in a single random allocation, and $\Delta_\varepsilon(1,1)$ is pure noise. 
This expression implies that the estimation error of $\hat{\tau}^{\DIF}$ is a sum of the chance imbalance and random noise, and becomes large when imbalanced covariates are highly prognostic (i.e. large magnitude of $\beta_0$). Similarly, if we substitute the true outcome surface in \eqref{eq:IPW}, we can show that the estimation error of IPW is $\hat{\tau}^{\IPW}-\tau=\Delta_X(1/(1-\hat{e}),1/\hat{e})^T\beta_0+\Delta_\varepsilon(1/(1-\hat{e}),1/\hat{e})$. Intuitively, IPW controls for chance imbalance because we usually have $\Vert{\Delta_X(1/(1-\hat{e}),1/\hat{e})}\Vert<\Vert{\Delta_X(1,1)}\Vert$, which reduces the variation of the estimation error over repeated experiments. However, because ${\Delta_X(1/(1-\hat{e}),1/\hat{e})}$ is not zero, the estimation error remains sensitive to the magnitude of $\beta_0$. In contrast, because of the exact mean balance property of OW, we have $\Delta_X(\hat{e},1-\hat{e})=0$; consequently, substituting the true outcome surface in \eqref{eq:OWATE}, we can see that the estimation error of OW equals $\hat{\tau}^{\OW}-\tau=\Delta_\varepsilon(\hat{e},1-\hat{e})$, which is only noise and free of $\beta_0$. This simple example illustrates that, for each realized randomization, OW should have the smallest estimation error, which translates into larger efficiency in estimating $\tau$ over repeated experiments.  

%Of note, more flexible models such as machine learning methods are available for estimating the propensity score \cite{Colantuoni2015}. In observational studies, these models could be more helpful because they reduce the chance for model misspecification. However, in randomized trials, the true propensity score is known and the propensity score model is merely a ``working" model that is never misspecified. Thus, the benefit of more flexible propensity score models in randomized trials is negligible. More importantly, we employ the logistic propensity score model to obtain the overlap weights that achieve the exact mean balance on baseline covariates. 

For non-continuous outcomes, we also consider ratio estimands. For example, while the ATE is also known as the causal risk difference with binary outcomes, $\tau=\tau_{\RD}$. Two other standard estimands are the causal risk ratio (RR) and the causal odds ratio (OR) on the log scale, defined by
\begin{align}
\label{eq:ratio_estimands}
\tau_{\RR}=\log\left(\frac{\mu_{1}}{\mu_{0}}\right),~~~~~
\tau_{\COR}=\log\left\{\frac{\mu_{1}/(1-\mu_{1})}{\mu_{0}/(1-\mu_{0})}\right\}.
\end{align}
The OW estimator for risk ratio and odds ratio are $\hat{\tau}_{\RR}=\log\{\hat{\mu}_{1}/\hat{\mu}_{0}\}$, and $\hat{\tau}_{\COR}=\log\{\hat{\mu}_{1}/(1-\hat{\mu}_{1})\}/\{\hat{\mu}_{0}/(1-\hat{\mu}_{0})\}$, respectively, with $\hat{\mu}_1$, $\hat{\mu}_0$ defined in \eqref{eq:OWATE}.

% Let $\mu_{1}=E\{Y_{i}(1)\}$ and $\mu_{0}=E\{Y_{i}(0)\}$, we are interested in three types of estimands, risk difference, risk ratio and odds ratio,

% \begin{align}
% \label{eq:three_estimands}
% \textup{risk difference:}\delta_{1}&=\tau=\mu_{1}-\mu_{0}\\
% \textup{risk ratio:}\tau_{\RR}&=\log (\mu_{1}/\mu_{0})\\
% \textup{odds ratio:}\tau_{\COR}&=\log (\mu_{1}/(1-\mu_{1})-\log(\mu_{0}/(1-\mu_{0})
% \end{align}

% Below is the general procedure of using propensity score weighting for covariate adjustment in randomized trials. First, estimate the propensity score, e.g. using a logistic regression model. Then, choose an $h(x)$ (equivalently a weighting scheme) and calculate the weights based on the estimated propensity score. Lastly, estimate the ATE using an unbiased estimator, e.g. the Hajek estimator:
% As shown before, as long as $h(x)$ is a function of the propensity score, the corresponding balancing weights will lead to an unbiased estimate of ATE. However, notable difference exists in the variance estimation, which is the key consideration in covariate adjustment in RCT. In particular, among all the balancing weights, the overlap weights, $\{w_1(x)=1-e(x), w_0(x)=e(x)\}$, resulting from the tilting function $h=e(x)(1-e(x))$, minimizes the asymptotic variance of the Hajek estimator $\hat{\tau}_w$ \cite{Crump2006, li2018balancing}. 

% \section{Variance reduction via overlap weighting}
\section{Efficiency Considerations and Variance Estimation}
\label{sec:theory}
In this section we demonstrate that in randomized trials the OW estimator leads to increased large-sample efficiency in estimating the treatment effect compared to the unadjusted estimator. We further propose a consistent variance estimator for the OW estimator of both the additive and ratio estimands. 

\subsection{Continuous Outcomes}\label{sec:cont}
%For continuous outcomes, the target estimand, $\tau$, is usually defined on the additive scale. 
Tsiatis et al. \cite{tsiatis2008covariate} show that the family of regular and asymptotically linear estimators for the additive estimand $\tau$ is
\begin{equation}
\label{eq:influence_function}
\mathcal{I}: \frac{1}{N}\sum_{i=1}^{N}
\left\{\frac{Z_{i}Y_{i}}{r}-\frac{(1-Z_{i})Y_{i}}{1-r}-\frac{Z_{i}-r}{r(1-r)}\left\{r
g_{0}(X_{i})+(1-r)g_{1}(X_{i})\right\}\right\}+o_{p}(N^{-1/2}),
\end{equation}
where $r$ is the randomization probability, and $g_{0}(X_{i})$, $g_{1}(X_{i})$ are scalar functions of the baseline covariates $X_{i}$.  
Several commonly used estimators for the treatment effect are members of the family $\mathcal{I}$, with different specifications of $g_{0}(X_{i})$, $g_{1}(X_{i})$. For example, setting $g_{0}(X_{i})=g_{1}(X_{i})=0$, we obtain the unadjusted estimator $\hat{\tau}^{\DIF}$. Setting $g_{0}(X_{i})=g_{1}(X_{i})=E(Y_{i}|X_{i})$, we obtain the ``ANCOVA I'' estimator in Yang and Tsiatis, \cite{yang2001efficiency} which is the least-squares solution of the coefficient of $Z_i$ in a linear regression of $Y_{i}$ on $Z_{i}$ and $X_{i}$.  %$g_{0}(X_{i})$, $g_{1}(X_{i})$ to be  projections of $Y_i$ into the space of $X_i$. 
Further, setting $g_{0}(X_{i})=E(Y_{i}|Z_{i}=0,X_{i})$ and $g_{1}(X_{i})=E(Y_{i}|Z_{i}=1,X_{i})$, we obtain the ``ANCOVA II'' estimator, \cite{yang2001efficiency,tsiatis2008covariate,lin2013agnostic} which is the least-squares solution of the coefficient of $Z_i$ in a linear regression of $Y_{i}$ on $Z_{i},X_{i}$ and their interaction terms. This estimator achieves the semiparametric variance lower bound within the family $\mathcal{I}$, when the conditional mean functions $g_{0}(X_{i})$
and $g_{1}(X_{i})$ are correctly specified in the ANCOVA model.\cite{robins1994estimation,leon2003semiparametric} Another member of $\mathcal{I}$ is the target maximum likelihood estimator, \cite{Moore2009,Moore2011,Colantuoni2015} which is asymptotic efficient under correct outcome model specification. The IPW estimator $\hat{\tau}^{\IPW}$ is also a member of $\mathcal{I}$. Specifically, Shen et al. \cite{shen2014inverse} showed that if the logistic model \eqref{eq:ps_logistic} is used to estimate the propensity score $\hat{e}_i$, then the IPW estimator is asymptotically equivalent to the ``ANCOVA II" estimator and becomes semiparametric efficient if the true $g_{0}(X_{i})$ and $g_{1}(X_{i})$ are linear functions of $X_i$.

In the following Proposition we show that the OW estimator is also a member of $\mathcal{I}$ and is asymptotically efficient under the linearity assumption. The proof of Proposition \ref{thm1} is provided in Web Appendix A.
%(Proposition \ref{thm1}), and then generalize the result to the family of balancing weights (Proposition \ref{prop:semi_bw}). \comment{FL: to be continued}
\begin{proposition}
(Asymptotic efficiency of overlap weighting)\label{thm1}\\
(a) If the propensity score is estimated by a parametric model $e(X;\theta)$ with parameters $\theta$ that satisfies a set of mild regularity conditions (specified in Web Appendix A), then $\hat{\tau}^{\OW}$ belongs to the class of estimators $\mathcal{I}$.\\
(b) Suppose $X^{1}$ and $X^{2}$ are two nested sets of baseline covariates with $X^{2}=(X^{1},X^{\ast 1})$, and $e(X^{1};\theta_{1})$, $e(X^{2};\theta_{2})$ are nested smooth parametric models. Write $\hat{\tau}^{\OW}_{1}$ and $\hat{\tau}^{\OW}_{2}$ as two OW estimators with the weights defined through $e(X^{1};\hat{\theta}_{1})$ and $e(X^{2};\hat{\theta}_{2})$, respectively. Then the asymptotic variance of $\hat{\tau}^{\OW}_{2}$ is no larger than that of $\hat{\tau}^{\OW}_{1}$.\\
(c) If the propensity score is estimated from the logistic regression \eqref{eq:ps_logistic}, then $\hat{\tau}^{\OW}$ is asymptotically equivalent to the ``ANCOVA II" estimator, and becomes semiparametric efficient as long as the true  $E(Y_{i}|X_{i},Z_{i}=1)$ and $E(Y_{i}|X_{i},Z_{i}=0)$ are linear in $X_i$.
\end{proposition}

Proposition \ref{thm1} summarizes the large-sample properties of the OW estimator in randomized trials, extending those demonstrated for IPW in Shen et al. \cite{shen2014inverse} In particular, adjusting for the baseline covariates using OW does not adversely affect efficiency in large samples than without adjustment. Further, the asymptotic equivalence between $\hat{\tau}^{\OW}$ and the ``ANCOVA II" estimator indicates that OW becomes fully semiparametric efficient when the conditional outcome surface is a linear function of the covariates adjusted in the logistic propensity score model. In the special case where the randomization probability $r=1/2$, we show in Web Appendix B that the limit of the large-sample variance of $\hat{\tau}^{\OW}$ is
\begin{align}
\label{eq:Rsqaure}
% N\cdot\textup{Var}(\hat{\tau}^{\OW})&\rightarrow (1-R^{2})N\textup{Var}(\hat{\tau}^{\DIF})\rightarrow4\left[\textup{Var}\left\{(2Z_i-1)Y_i\right\}-\left\{E(Y_i)\right\}^2-R^2 \textup{Var}(Y_i)\right],\\
\lim_{N\rightarrow\infty}N\textup{Var}(\hat{\tau}^{\OW})=(1-R^{2}_{\tilde{Y}\sim X})\lim_{N\rightarrow\infty} N\textup{Var}(\hat{\tau}^{\DIF})=4  (1-R^{2}_{\tilde{Y}\sim X})\textup{Var}(\tilde{Y}_{i}) ,
\end{align}
where $\tilde{Y}_{i}=Z_{i}(Y_{i}-\mu_{1})+(1-Z_{i})(Y_{i}-\mu_{0})$ is the mean-centered outcome and $R^{2}_{\tilde{Y}\sim X}$ measures the proportion of explained variance after regressing $\tilde{Y}_{i}$ on $X_{i}$. Similar definition of $R$-squared was also used elsewhere when demonstrating efficiency gain with covariate adjustment.\cite{Moore2009, Moore2011, wang2019analysis} The amount of variance reduction is also a direct result from the asymptotic equivalence between the OW, IPW, and ``ANCOVA II'' estimators. %For those estimators, $R^{2}_{\tilde{Y}\sim X}$ determines the efficiency gain of the OW estimator compared with the unadjusted estimator.
%corresponds to the ``$R^{2}_{Y-Z\sim X}$'' in Wang et al. \cite{wang2019analysis}. A similar $R^2$ is also defined in Moore and van der Laan \cite{Moore2009} and Moore et at. \cite{Moore2011} for demonstrating the efficiency gain using the target maximum likelihood estimation. 
Equation \eqref{eq:Rsqaure} shows that incorporating additional covariates into the propensity score model will not reduce the asymptotic efficiency because $R^{2}_{\tilde{Y}\sim X}$ is non-decreasing when more covariates are considered. Although adding covariates does not hurt the asymptotic efficiency, in practice we recommend incorporating the covariates that exhibit baseline imbalance and that have large predictive power for the outcome.\cite{williamson2014variance}

Perhaps more interestingly, the results in Proposition \ref{thm1} apply more broadly to the family of balancing weights estimators, formalized in the following Proposition. The proof of Proposition \ref{prop:semi_bw} is presented in Web Appendix A.
\begin{proposition}(Extension to balancing weights) \label{prop:semi_bw}\\
Proposition \ref{thm1} holds for the general family of estimators \eqref{eq:sampleWATE} using balancing weights defined in \eqref{def:bw}, as long as the tilting function $h(X)$ is a ``smooth'' function of the propensity score, where ``smooth'' is defined by  satisfying a set of mild regularity conditions (specified in details in Web Appendix A).
%following properties\comment{SZ: might be too specific?}. If we express $h(x)$ and balancing weights $w_{1}(x),w_{0}(x)$ as a function $\theta$, $h(X_{i};\theta),w_{1}(X_{i};\theta),w_{0}(X_{i};\theta)$. We assume $h(x;\theta),w_{1}(x;\theta),w_{0}(x;\theta)$ satisfy:
% \begin{itemize}
%     \item[(i)] (Nonzero tilting function) There exists $\varepsilon>0$ such that $P(h(X_{i};\theta_{0})>\varepsilon)=1$, where $\theta_{0}$ is the true value.
%     \item[(ii)] (Smoothness) the first and second order derivative of balancing weights with respect to the propensity score $\frac{d}{d e} w_{1}(X_{i};\theta)\},\frac{d}{d e} w_{0}(X_{i};\theta)$,$\frac{d^{2}}{d e^{2}} w_{1}(X_{i};\theta)\},\frac{d^{2} }{d e^{2}} w_{0}(X_{i};\theta)$ exists and are continuous in $e$.
%     \item[(iv)] (Bounded derivative in the neighborhood of $\theta_{0}$) For the true value $\theta_{0}$, there exists $c>0$ and $M_{1}>0,M_{2}>0$  such that 
%     \begin{align*}
%     \left|\frac{d}{d e} w_{0}(X_{i};\theta_{0})\right|\leq M_{1},&\left|\frac{d}{d e} w_{0}(X_{i};\theta_{0})\right|\leq M_{1}\\
%         \left| \frac{d^{2}}{d e^{2}} w_{0}(X_{i};\theta) \right|\leq M_{2},&\left| \frac{d^{2}}{d e^{2}} w_{1}(X_{i};\theta) \right|\leq M_{2},
%     \end{align*}
%     almost surely for $\theta$ in the neighborhood of $\theta_{0}$, i.e. $\theta \in \{\theta|\left|\left|\theta-\theta_{0}\right|\right|_{1}\leq c\}$.
% \end{itemize}
\end{proposition}

% \comment{need h(x) as a smooth function of e(x)}

\subsection{Binary Outcomes}
For binary outcomes, the target estimand could be the causal risk difference, risk ratio and odds ratio, denoted as $\tau_{\RD}$, $\tau_{\RR}$ and $\tau_{\COR}$, respectively. The discussions in Section \ref{sec:cont} directly apply to the estimation of the additive estimand, $\tau_{\RD}$. When estimating the ratio estimands, one should proceed with caution in interpreting regression parameters for the ANCOVA-type generalized linear models due to the potential non-collapsibility issue. Additionally, it is well-known that the log-binomial model frequently fails to converge with a number of covariates, and therefore one may have to resort to less efficient regression methods such as the modified Poisson regression.\cite{Zou2004} Williamson et al. \cite{williamson2014variance} showed that IPW can be used to adjust for baseline covariates without changing the interpretation of the marginal treatment effect estimands, $\tau_{\RR}$ and $\tau_{\COR}$. Because of the asymptotic equivalence between the IPW and OW estimators (Proposition \ref{thm1}), OW shares the advantages of IPW in improving the asymptotic efficiency over the unadjusted estimators for risk ratio and odds ratio without compromising the interpretation of the marginal estimands. In addition, due to its ability to remove all chance imbalance associated with $X_i$, OW is likely to give higher efficiency than IPW in finite samples, which we will demonstrate in Section \ref{sec:simulations}.

\subsection{Variance Estimation}\label{sec:variance}
To estimate the variance of the propensity score estimators, it is important to incorporate the uncertainty in estimating the propensity scores. \cite{lunceford2004stratification} Failing to do so leads to conservative variance estimates of the weighting estimator and therefore reduces power of the Wald test for treatment effect. \cite{williamson2014variance} Below we use the M-estimation theory \cite{tsiatis2007semiparametricbook} to derive a consistent variance estimator for OW. Specifically, we cast $\hat{\mu}_{1},\hat{\mu}_{0}$ in equation \eqref{eq:OWATE}, and $\hat{\theta}$ in the logistic model \eqref{eq:ps_logistic} as the solutions $\hat{\lambda}=(\hat{\mu}_{1},\hat{\mu}_{0},\hat{\theta}^T)^T$ to the following joint estimation equations $\sumi U_i=\sum_{i=1}^{N}U(Y_{i},X_{i},Z_{i};\hat{\lambda})=0$, where
\begin{equation}
\label{eq:estimation_equations}
\sum_{i=1}^{n}U(Y_{i},X_{i},Z_{i},\lambda)=\sum_{i=1}^{N}\begin{bmatrix}
Z_{i}(Y_{i}-\mu_{1})(1-e_{i})\\
(1-Z_{i})(Y_{i}-\mu_{0})e_{i}\\
\tilde{X}_{i}(Z_{i}-e_{i})\\
\end{bmatrix}=0,
\end{equation}
where $\tilde{X}_{i}=(1,X_{i}^{T})^{T}$ is the augmented covariates with an intercept. Here, the first two rows represent the estimating functions for $\hat{\mu}_{1}$ and $\hat{\mu}_{0}$ and the last rows are the score functions of the logistic model with an intercept and main effects of $X_{i}$. If $X_{i}$ is of $p$ dimensions, equation \eqref{eq:estimation_equations} involves $p+3$ scalar estimating equations for $p+3$ parameters. %Based on the estimation equations in \eqref{eq:estimation_equations}, we can also derive an empirical sandwich estimator for the asymptotic variance of  $\hat{\tau}^{\OW}$, which accounts for the uncertainty in estimating the propensity scores. We use bold font to denote matrix and vector. 
Let $A=-E(\partial U_i/\partial \lambda)^{T}$,$B=E(U_iU_i^{T})$, the asymptotic covariance matrix for $\hat{\lambda}$ can be written as $N^{-1}A^{-1}BA^{-T}$. Extracting the covariance matrix for the first two components in $\hat{\lambda}$,
% where the matrices are of the form
% \begin{equation*}
%     \mathbf{A}=\begin{bmatrix}
%     a_{11}&0&\mathbf{a}_{13}\\
%     0&a_{22}&\mathbf{a}_{23}\\
%     \textbf{0}_{p\times 1}&\textbf{0}_{p\times 1}&\mathbf{a}_{33}
%     \end{bmatrix},~~~~~~~\mathbf{B}=\begin{bmatrix}
%  b_{11}&0&\mathbf{b}_{13}\\
%     0&b_{22}&\mathbf{b}_{23}\\
% \mathbf{b}_{13}^{T}&\mathbf{b}_{23}^{T}&\mathbf{b}_{33}
%     \end{bmatrix}
% \end{equation*},
% with elements in the matrices defined in Web Appendix xx.
we can show that, as $N$ goes to infinity,
\begin{align}
\label{eq:asy_var_ow}
\sqrt{N} \begin{bmatrix}
\hat{\mu}_{1}-\mu_{1}\\
\hat{\mu}_{0}-\mu_{0}
\end{bmatrix}&\rightarrow \mathcal{N}\left\{\textbf{0},\begin{bmatrix}
\Sigma_{11},\Sigma_{12}\\
\Sigma_{21},\Sigma_{22}
\end{bmatrix}\right\},
\end{align}
where the covariance matrix is defined as the corresponding elements in $A^{-1}BA^{-T}$,
\begin{align}
\Sigma_{11}=[A^{-1}BA^{-T}]_{1,1},\Sigma_{22}=[A^{-1}BA^{-T}]_{2,2},\Sigma_{12}=\Sigma_{21}^T=[A^{-1}BA^{-T}]_{1,2}. \label{eq:covariance_value}
\end{align}
where $[A^{-1}BA^{-T}]_{j,k}$ denotes the $(j,k)$th element of the matrix $A^{-1}BA^{-T}$. Using the delta method, we can obtain the asymptotic variance of $\hat{\tau}_{\RD}^{\OW}$, $\hat{\tau}_{\RR}^{\OW}$, $\hat{\tau}_{\COR}^{\OW}$ as a function of $\Sigma_{11},\Sigma_{22},\Sigma_{12}$.
% \begin{gather}
% \label{eq:asymp_var_three_estimands_1}
% \textup{Var}(\hat{\tau}_{\RD}^{\OW})=\frac{1}{N}\left(\Sigma_{11}+\Sigma_{22}-2\Sigma_{12}\right),\\
% \label{eq:asymp_var_three_estimands_2}
% \textup{Var}(\hat{\tau}_{\RR}^{\OW})=\frac{1}{N}\left(\frac{\Sigma_{11}}{\mu_{1}^{2}}+\frac{\Sigma_{22}}{\mu_{0}^{2}}-\frac{2\Sigma_{12}}{\mu_{1}\mu_{0}}\right), \\ 
% \label{eq:asymp_var_three_estimands_3}
% \textup{Var}(\hat{\tau}_{\COR}^{\OW})=\frac{1}{N}\left\{\frac{\Sigma_{11}}{\mu_{1}^{2}(1-\mu_{1})^{2} }+\frac{\Sigma_{22}}{\mu_{0}^{2}(1-\mu_{0})^{2} }-\frac{2\Sigma_{12}}{\mu_{1}(1-\mu_{1})\mu_{0}(1-\mu_{0})}\right\}.
% \end{gather}
Consistent plug-in estimators can then be obtained by estimating the expectations in the ``sandwich'' matrix $A^{-1}BA^{-T}$ by their corresponding sample averages. We summarize the variance estimators for $\hat{\tau}_{\RD}^{\OW},\hat{\tau}_{\RR}^{\OW},\hat{\tau}_{\COR}^{\OW}$ in the following general equations,
\begin{gather}\label{eq:varBin}
\textup{Var}(\hat{\tau}^{\OW})=\frac{1}{N}\left[\hat{V}^{\DIF}-\hat{v}_{
1}^{T}
\left\{\frac{1}{N}\sum_{i=1}^{N}\hat{e}_{i}(1-\hat{e}_{i})\tilde{X}_{i}^{T}\tilde{X}_{i}\right\}^{-1}
(2\hat{v}_{1}-\hat{v}_{2})\right],
\end{gather}
where
\begin{gather*}
\hat{V}^{\DIF}=\left\{\frac{1}{N}\sum_{i=1}^{N}\hat{e}_{i}(1-\hat{e}_{i})\right\}^{-1}
\left(\frac{\hat{E}_{1}^{2} }{N_1}\sum_{i=1}^{N}Z_i\hat{e}_{i}(1-\hat{e}_{i})^{2}(Y_{i}-\hat{\mu}_{1})^{2}+\frac{\hat{E}_{0}^{2}}{N_0}\sum_{i=1}^{N}(1-Z_i)\hat{e}_{i}^{2}(1-\hat{e}_{i})(Y_{i}-\hat{\mu}_{0})^{2}\right),\\
\hat{v}_{1}=\left\{\frac{1}{N}\sum_{i=1}^{N}\hat{e}_{i}(1-\hat{e}_{i})\right\}^{-1}\left(\frac{\hat{E}_{1}}{N_1}\sum_{i=1}^{N}Z_i\hat{e}_{i}^{2}(1-\hat{e}_{i})(Y_{i}-\hat{\mu}_{1})^{2}\tilde{X}_{i}+\frac{\hat{E}_{0}}{N_0}\sum_{i=1}^{N}(1-Z_i)\hat{e}_{i}(1-\hat{e}_{i})^{2}(Y_{i}-\hat{\mu}_{0})^{2}\tilde{X}_{i}\right),\\
\hat{v}_{2}=\left\{\frac{1}{N}\sum_{i=1}^{N}\hat{e}_{i}(1-\hat{e}_{i})\right\}^{-1}\left(\frac{\hat{E}_{1}}{N_1}\sum_{i=1}^{N}Z_i\hat{e}_{i}(1-\hat{e}_{i})^{2}(Y_{i}-\hat{\mu}_{1})^{2}\tilde{X}_{i}+\frac{\hat{E}_{0}}{N_0}\sum_{i=1}^{N}(1-Z_i)\hat{e}_{i}^{2}(1-\hat{e}_{i})(Y_{i}-\hat{\mu}_{0})^{2}\tilde{X}_{i}\right),\\
\end{gather*}
and $\hat{E}_{k}$ depends on the choice of estimands. For $\hat{\tau}^{\OW}_{\RD}$, we have $\hat{E}_{k}=1$; for $\hat{\tau}^{\OW}_{\RR}$, we set $\hat{E}_{k}=\hat{\mu}_{k}^{-1}$; for $\hat{\tau}^{\OW}_{\COR}$, we use $\hat{E}_{k}=\hat{\mu}_{k}^{-1}(1-\hat{\mu}_{k})^{-1}$ with $k=0,1$. Detailed derivation of the asymptotic variance and its consistent estimator can be found in Web Appendix B. These variance calculations are implemented in the R package \textbf{PSweight}.\cite{zhou2020psweight}

\section{Simulation Studies} \label{sec:simulations}
We carry out extensive simulations to investigate the finite-sample operating characteristics of OW relative to IPW, direct regression adjustment and an augmented estimator that combined IPW and outcome regression. The main purpose of the simulation study is to empirically (i) illustrate that OW leads to marked finite-sample efficiency gain compared with IPW in estimating the treatment effect, and (ii) validate the sandwich variance estimator of OW developed in Section \ref{sec:variance}. Below we focus on the simulations with continuous outcomes. We have also conducted extensive simulations with binary outcomes, the details of which are presented in Web Appendix D. 

\subsection{Simulation Design}\label{sec:cont_simu}
We generate $p=10$ baseline covariates from the standard normal distribution, $X_{ij}{\sim} \mathcal{N}(0,1)$, $j=1,2,\cdots,p$. Fixing the randomization probability $r$, the treatment indicator is randomly generated from a Bernoulli distribution, $Z_{i}\sim\textup{Bern}(r)$. Given the baseline covariates $X_i=(X_{i1},\ldots,X_{ip})^T$, we generate the potential outcomes from the following linear model (model 1): for $z=0,1$,
\begin{gather}
\label{eq:cont_simu}
Y_{i}(z){\sim}  \mathcal{N}(z\alpha+X_{i}^T\beta_{0}+zX_{i}^T\beta_{1},\sigma_{y}^{2}), \quad i=1,2,\cdots,N
\end{gather}
where $\alpha$ is the main effect of the treatment, and $\beta_{0}, \beta_{1}$ are the effects of the covariates and treatment-by-covariate interactions. The observed outcome is set to be $Y_i=Y_i(Z_i)=Z_iY_i(1)+(1-Z_i)Y_i(0)$. In our data generating process, because the baseline covariates have mean zero, the true average treatment effect on the additive scale $\tau=\alpha$. 
For simplicity, we fix $\tau=0$ and choose $\beta_{0}=b_{0}\times (1,1,2,2,4,4,8,8,16,16)^{T}$, $\beta_{1}=b_{1}\times (1,1,1,1,1,1,1,1,1,1)^{T}$. We specify the residual variance $\sigma_{y}^{2}=2$, and choose the multiplication factor $b_0$ so that the signal-to-noise ratio (due to the main effects) is 1, namely, $\sum_{i=1}^{p}\beta_{0i}^{2}/\sigma_{y}^{2}=1$. This specification mimics a scenario where the baseline covariates can explain up to $50\%$ of the variation in the outcome. We also assign different importance to each covariates. For example, the last two covariates, $X_{9}$, $X_{10}$, explain the majority of the variation, mimicking the scenario that one may have access to only a few strong prognostic risk factors. We additionally vary the value of $b_{1}\in\{0,0.25,0.5,0.75\}$ to control the strength of treatment-by-covariate interactions. A larger value of $b_{1}$ indicates a higher level of treatment effect heterogeneity so that the baseline covariates are more strongly associated with the individual-level treatment contrast, $Y_i(1)-Y_i(0)$. For brevity, we present the results with $b_{1}=0.25,0.5$ to the Web Appendix and focus here on the scenarios with homogeneous treatment effect ($b_{1}=0$) and with the strongest effect heterogeneity ($b_{1}=0.75$). For the randomization probability $r$, we consider two values: $r=0.5$ represents a balanced design with one-to-one randomization, and $r=0.7$ an unbalanced assignment where more patients are randomized to the treatment arm. We also vary the total sample sizes $N$ from $50$ to $500$, with $50$ and $500$ mimicking a small and large sample scenario, respectively. 
% Findings under other scenarios with $r=0.6$ are similar to those with $r=0.7$ and are omitted for brevity.
%comparable to that in Raad et al. \cite{Raad2020}. 
% The results for $b_1=0.25,0.5,0.75$ are similar, and we mostly focus on results for $b_1=0$ and $b_1=0.75$. 

In each simulation scenario, we compare several different estimators for ATE, including the unadjusted estimator $\hat{\tau}^{\DIF}$ (UNADJ), the IPW estimator $\hat{\tau}^{\IPW}$, the estimator based on linear regression $\hat{\tau}^{\LR}$ (LR), and the OW estimator $\hat{\tau}^{\OW}$. For the IPW and OW estimators, we estimate the propensity score by logistic regression including all baseline covariates as linear terms, and the final estimator is given by the H\'{a}jek-type estimator \eqref{eq:sampleWATE} using the corresponding weights. For the LR estimator, we fit the correctly specified outcome model \eqref{eq:cont_simu} (model 1). In addition, we also consider an augmented IPW (AIPW) estimator that augments IPW with an outcome regression, \cite{lunceford2004stratification} which is also a member of the class $\mathcal{I}$:
\begin{equation}
\label{eq:aipw}
\hat{\tau}^{\AIPW}=\hat{\mu}_1^{\AIPW}-\hat{\mu}_0^{\AIPW}=\frac{1}{N}\sum_{i=1}^{N}
\left\{\frac{Z_{i}Y_{i}}{\hat{e}_i}-\frac{(Z_{i}-\hat{e}_i)\hat{\mu}_1(X_i)}{\hat{e}_i}\right\}-\left\{ \frac{(1-Z_{i})Y_{i}}{1-\hat{e}_i}+\frac{(Z_{i}-\hat{e}_i)\hat{\mu}_0(X_i)}{1-\hat{e}_i}\right\},
\end{equation}
where $\hat{\mu}_z(X_i)=\hat{E}[Y_i|X_i,Z_i=z]$ is the prediction from the outcome regression. In the context of observational studies, such an estimator is also known as the doubly-robust estimator. Because AIPW hybrids propensity score weighting and outcome regression, it does not retain the objectivity of the former. Nonetheless, the AIPW estimator is often perceived as an improved version of IPW \cite{Bang2005}; therefore, we also compare it in the simulations to understand its operating characteristics in randomized trials. 

% In fact, the AIPW estimator is also a member of the class $\mathcal{I}$ in equation \eqref{eq:influence_function} and is semiparametric efficient when both the propensity score and the outcome models are correctly specified. 

% \comment{FL: add a small paragraph AIPW here.} We also compare with the popular augmented IPW (AIPW) estimator \cite{lunceford2004stratification}, also known as the double-robust estimator \cite{bang2005}, which augments the IPW estimator by an outcome regression model:
%In observational studies, the AIPW estimator, also known as the double-robust estimator \cite{bang2005}, 

For each scenario, we simulate 2000 replicates, and calculate the bias, Monte Carlo variance and mean squared error for each estimator of $\tau$. Across all scenarios, as expected we find that the bias of all estimators is negligible, and thus the Monte Carlo variance and the mean squared error are almost identical. For this reason, we focus on reporting the efficiency comparisons using the Monte Carlo variance. We define the \emph{relative efficiency} of an estimator as the ratio between the Monte Carlo variance of that estimator and that of the unadjusted estimator. Relative efficiency larger than one indicates that estimator is more efficient than the unadjusted estimator. We also examine the empirical coverage rate of the associated 95\% normality-based confidence intervals. Specifically, the confidence interval of $\hat{\tau}^{\LR}$, $\hat{\tau}^{\IPW}$, and  $\hat{\tau}^{\OW}$ is constructed based on the Huber-White estimator in Lin,\cite{lin2013agnostic} the sandwich estimator in Williamson et al.,\cite{williamson2014variance} and the sandwich estimator developed in Section \ref{sec:variance}, respectively. The confidence interval of $\hat{\tau}^{\AIPW}$ is the based on the sandwich variance derived based on the M-estimation theory; the details are presented in Web Appendix C.

To explore the performance of the estimators under model misspecification, we repeat the simulations by replacing the potential outcome generating process with the following model (model 2)
\begin{align}
\label{eq:misspecified}
Y_{i}(z){\sim}  \mathcal{N}(z\alpha+X_{i}^T\beta_{0}+zX_i^T\beta_1+X_{i,\textup{int}}^T\gamma,\sigma_{y}^{2}),
\end{align}
where $X_{i,\textup{int}}=(X_{i1}X_{i2},X_{i2}X_{i3},\cdots,X_{ip-1}X_{ip})$ represents $p-1$ interactions between pairs of covariates with consecutive indices and $\gamma=\sqrt{\sigma_{y}^{2}/p}\times (1,1,\cdots,1)^{T}$ represents the strength of this interaction effect. The LR estimator omitting these additional interactions is thus considered as misspecified. For IPW and OW, the propensity score model is technically correctly specified (because the true randomization probability is a constant) even though it does not adjust for the interaction term $X_{i,\textup{int}}$. The AIPW estimator similarly omits $X_{i,\textup{int}}$ in both the propensity score and outcome models. With a slight abuse of terminology, we refer to this scenario as ``model misspecification."

\subsection{Results on Efficiency of Point Estimators}

\begin{figure}[htbp]
\centering
\includegraphics[width=1\textwidth]{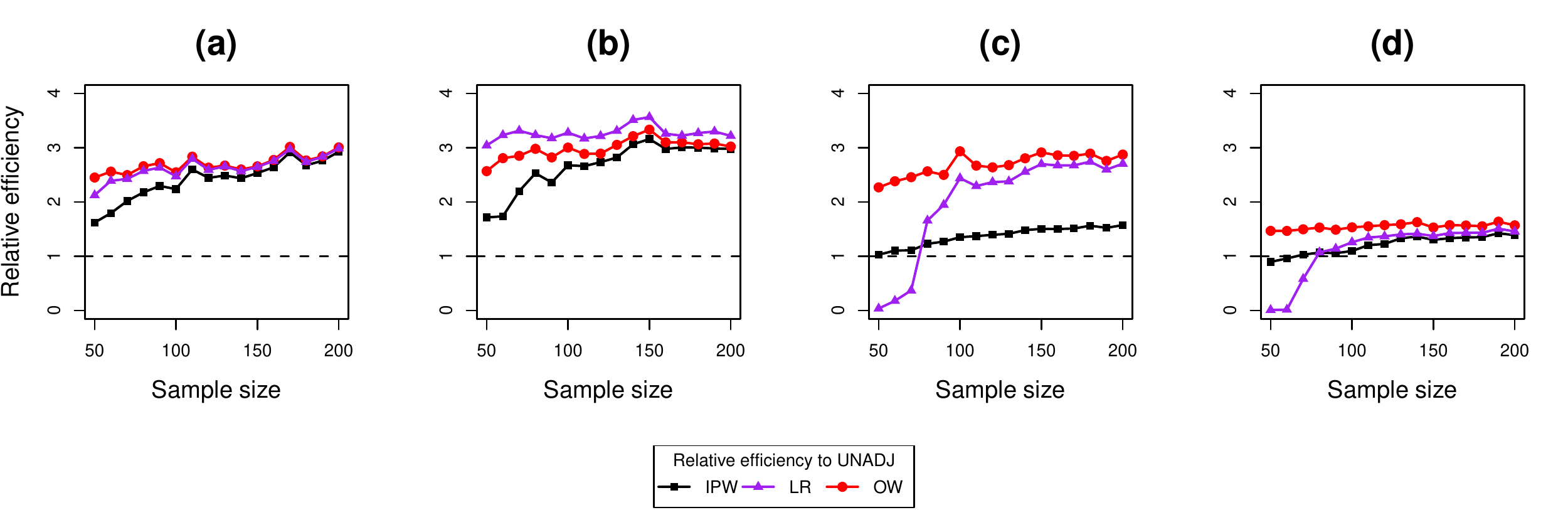}
\caption{\label{fig:performance_cont_mcsd} The relative efficiency of $\hat{\tau}^{\IPW}$, $\hat{\tau}^{\AIPW}$, $\hat{\tau}^{\LR}$ and $\hat{\tau}^{\OW}$ relative to $\hat{\tau}^{\DIF}$ for estimating ATE when (a) $r=0.5$, $b_{1}=0$ and the outcome model is correctly specified, (b) $r=0.5,b_{1}=0.75$ and the outcome model is correctly specified, (c) $r=0.7,b_{1}=0$ and the outcome model is correctly specified, (d) $r=0.7,b_{1}=0$ and the outcome model is misspecified. A larger value of relative efficiency corresponds to a more efficient estimator.}
\end{figure}

Figure \ref{fig:performance_cont_mcsd} presents the relative efficiency of the different estimators in four typical scenarios. For a more clear presentation, we omit the results for $\hat{\tau}^{\AIPW}$ as they become indistinguishable from the results for $\hat{\tau}^{\LR}$ in these scenarios. Below, we discuss in order the relative efficiency results when the outcomes are generated under model 1 (panel (a) to (c)) and model 2 (panel (d)). 

Panel (a) to (c) correspond to scenarios when the outcomes are simulated from model 1. When $r=0.5$ and there is no treatment effect heterogeneity (panel (a)), it is evident that $\hat{\tau}^{\IPW},\hat{\tau}^{\LR}$, and $\hat{\tau}^{\OW}$ are consistently more efficient than the unadjusted estimator, and the relative efficiency increases with a larger sample size. However, when the sample size is no larger than $100$, OW leads to higher efficiency compared to LR and IPW, with IPW being the least efficient among the adjusted estimators. With a strong treatment effect heterogeneity $b_{1}=0.75$ (panel (b)), $\hat{\tau}^{\LR}$ becomes slightly more efficient than $\hat{\tau}^{\OW}$; this is expected as the true outcome model is used and the design is balanced. The efficiency advantage decreases for $\hat{\tau}^{\LR}$ and as $b_1$ moves closer to zero (see Web Table 1). On the other hand, $\hat{\tau}^{\OW}$ becomes more efficient than $\hat{\tau}^{\LR}$ when the randomization probability deviates from $0.5$. For instance, in panel (c), with $r=0.7$ and $N=50$, $\hat{\tau}^{\LR}$ becomes even less efficient than the unadjusted estimator, while OW demonstrates substantial efficiency gain over the unadjusted estimator. The deteriorating performance of $\hat{\tau}^{\LR}$ under $r=0.7$ also supports the findings in Freedman. \cite{Freedman2008} These results show that the relative performance between LR and OW is affected by the degree of treatment effect heterogeneity and the randomization probability. In the scenarios with a small degree of effect heterogeneity and/or with unbalanced design, OW tends to be more efficient than LR. 

Overall, OW is generally comparable to LR with a correctly specified outcome model, both outperforming IPW. But OW becomes more efficient than LR when the outcome model is incorrectly specified. Namely, when the outcomes are generated from model 2, $\hat{\tau}^{\OW}$ becomes the most efficient even if the propensity model omits important interaction terms in the true outcome model, as in panel (d) of Figure \ref{fig:performance_cont_mcsd}. The fact that LR and AIPW have almost identical finite-sample efficiency further confirms that the regression component dominates the AIPW estimator in randomized trials. Throughout, $\hat{\tau}^{\OW}$ is consistently more efficient than $\hat{\tau}^{\IPW}$, regardless of sample size, randomization probability and the degree of treatment effect heterogeneity. % These patterns persist when the sample size increases to $N=500$, although there the differences between methods become smaller as a result of Proposition \ref{thm1}. 
When the sample size increases to $N=500$, the differences between methods become smaller as a result of Proposition \ref{thm1}. Additional results on relative efficiency are also provided in Table \ref{tab:cont_simu_results} and Web Appendix Table 1.

\subsection{Results on Variance and Interval Estimators}

Table \ref{tab:cont_simu_results} summarizes the accuracy of the estimated variance and the empirical coverage rate of each interval estimator in four scenarios that match Figure \ref{fig:performance_cont_mcsd}. The former is measured by the ratio between the average estimated variance and the Monte Carlo variance of each estimator, and a ratio close to $1$ indicates adequate performance. In general, we find that estimated variance is close to the truth for both IPW and OW, but less so for the LR and AIPW estimator, especially with small sample sizes such as $N=50$ or $100$. Specifically, when the outcomes are generated from model 1, the sandwich variances of IPW and OW estimators usually adequately quantify the uncertainty, even when the sample size is as small as $N=50$. In the same settings, the Huber-White variance estimator for LR sometimes \emph{substantially underestimates} the true variance, leading to under-coverage of the interval estimator. Also, in the case where LR has a slight efficiency advantage ($b_{1}=0.75$), the coverage of LR is only around 70\% even when the true linear regression model is estimated. This result shows that the Huber-White sandwich variance, although known to be robust to heteroscedasticity in large samples, could be severely biased towards zero in finite samples when there is treatment effect heterogeneity. Further, the sandwich variance of AIPW also frequently underestimates the true variance when $N=50$ and $100$. On the other hand, when the outcomes are generated from model 2 and the randomization probability is $r=0.7$, all variance estimators tend to underestimate the truth, and the coverage rate slightly deteriorates. However, the coverage of the IPW and OW estimators is still closer to nominal than LR and AIPW when $N=50$ and $100$.
% suggesting that the sandwich variance of the propensity score weighting estimators have more stable performance.

\begin{table}[htbp]
\centering
\caption{\label{tab:cont_simu_results} The relative efficiency of each estimator compared to the unadjusted estimator, the ratio between the average estimated variance over Monte Carlo variance (\{Est Var\}/\{MC Var\}), and 95\% coverage rate of IPW, LR, AIPW and OW estimators. The results are based on 2000 simulations with a continuous outcome. In the ``correct specification" scenario, data are generated from model 1; in the "misspecification" scenario, data are generated from model 2. For each estimator, the same analysis approach is used throughout, regardless of the data generating model.}
\resizebox{1\textwidth}{!}{
\begin{tabular}{c@{\hskip 0.3in}cccc@{\hskip 0.3in}cccc@{\hskip 0.3in}cccc}
\hline
Sample size& \multicolumn{4}{c}{Relative efficiency}& \multicolumn{4}{c}{\{Est Var\}/\{MC Var\}}&\multicolumn{4}{c}{95\% Coverage}\\
$N$   & IPW & LR &AIPW & OW & IPW & LR &AIPW & OW  & IPW & LR &AIPW & OW \\ 
\hline
\multicolumn{13}{c}{$r=0.5$, $b_{1}=0$, correct specification}\\
50 & 1.621 & 2.126 & 2.042 & 2.451 & 1.001 & 0.866 & 0.668 & 1.343 & 0.936 & 0.933 & 0.885 & 0.967 \\ 
  100 & 2.238 & 2.475 & 2.399 & 2.548 & 0.898 & 0.961 & 0.799 & 1.116 & 0.938 & 0.944 & 0.914 & 0.955 \\ 
  200 & 2.927 & 2.987 & 2.984 & 3.007 & 0.951 & 0.996 & 0.927 & 1.051 & 0.946 & 0.949 & 0.938 & 0.956 \\ \smallskip
  500 & 2.985 & 3.004 & 2.995 & 3.006 & 0.963 & 0.987 & 0.959 & 1.000 & 0.944 & 0.949 & 0.942 & 0.952 \\
% \multicolumn{13}{c}{$r=0.5$, $b_{1}=0.25$, correct specification}\\
%   50 & 1.910 & 2.792 & 2.606 & 2.905 & 1.141 & 0.711 & 0.684 & 1.562 & 0.946 & 0.899 & 0.887 & 0.972 \\ 
%   100 & 2.968 & 3.575 & 3.481 & 3.489 & 0.988 & 0.811 & 0.896 & 1.295 & 0.954 & 0.925 & 0.928 & 0.968 \\ 
%   200 & 3.640 & 3.864 & 3.855 & 3.794 & 0.932 & 0.754 & 0.923 & 1.079 & 0.940 & 0.912 & 0.933 & 0.956 \\ \smallskip
%   500 & 3.801 & 3.814 & 3.814 & 3.791 & 0.947 & 0.735 & 0.940 & 0.992 & 0.945 & 0.907 & 0.945 & 0.950 \\ 
% \multicolumn{13}{c}{$r=0.5$, $b_{1}=0.5$, correct specification}\\
%   50 & 1.635 & 2.894 & 2.781 & 2.755 & 1.021 & 0.463 & 0.769 & 1.530 & 0.936 & 0.822 & 0.910 & 0.970 \\ 
%   100 & 3.084 & 3.917 & 3.835 & 3.546 & 0.984 & 0.510 & 0.977 & 1.291 & 0.942 & 0.840 & 0.944 & 0.968 \\ 
%   200 & 3.187 & 3.410 & 3.406 & 3.287 & 0.924 & 0.446 & 0.936 & 1.061 & 0.944 & 0.802 & 0.942 & 0.956 \\ \smallskip
%   500 & 3.730 & 3.809 & 3.810 & 3.717 & 1.037 & 0.477 & 1.049 & 1.085 & 0.957 & 0.818 & 0.960 & 0.962 \\ 
\multicolumn{13}{c}{$r=0.5$, $b_{1}=0.75$, correct specification}\\
  50 & 1.715 & 3.043 & 2.972 & 2.570 & 0.991 & 0.286 & 0.816 & 1.383 & 0.935 & 0.712 & 0.918 & 0.967 \\ 
  100 & 2.679 & 3.279 & 3.253 & 3.003 & 0.931 & 0.280 & 0.917 & 1.168 & 0.942 & 0.710 & 0.934 & 0.966 \\ 
  200 & 2.979 & 3.220 & 3.215 & 3.023 & 0.967 & 0.278 & 0.995 & 1.075 & 0.951 & 0.697 & 0.949 & 0.964 \\ \smallskip
  500 & 3.337 & 3.425 & 3.426 & 3.338 & 0.995 & 0.273 & 1.013 & 1.037 & 0.943 & 0.696 & 0.945 & 0.954 \\ 
% \multicolumn{13}{c}{$r=0.6$, $b_{1}=0$, correct specification}\\
%   50 & 1.415 & 1.686 & 1.605 & 2.418 & 1.041 & 0.745 & 0.617 & 1.377 & 0.938 & 0.913 & 0.883 & 0.959 \\ 
%   100 & 2.042 & 2.378 & 2.290 & 2.521 & 0.889 & 0.942 & 0.784 & 1.104 & 0.944 & 0.941 & 0.915 & 0.956 \\ 
%   200 & 2.777 & 2.926 & 2.896 & 2.981 & 0.987 & 1.027 & 0.947 & 1.078 & 0.949 & 0.950 & 0.940 & 0.953 \\ \smallskip
%   500 & 2.898 & 2.939 & 2.939 & 2.950 & 0.976 & 0.994 & 0.969 & 1.003 & 0.953 & 0.953 & 0.949 & 0.953 \\ 
\multicolumn{13}{c}{$r=0.7$, $b_{1}=0$, correct specification}\\
  50 & 1.056 & 0.036 & 0.036 & 2.270 & 1.060 & 0.014 & 0.026 & 1.184 & 0.938 & 0.779 & 0.816 & 0.931 \\ 
  100 & 1.825 & 2.439 & 2.311 & 2.935 & 0.914 & 0.858 & 0.717 & 1.039 & 0.946 & 0.921 & 0.897 & 0.923 \\ 
  200 & 2.474 & 2.706 & 2.679 & 2.874 & 0.971 & 0.931 & 0.857 & 0.963 & 0.948 & 0.944 & 0.927 & 0.935 \\ \smallskip
  500 & 2.641 & 2.743 & 2.738 & 2.809 & 0.922 & 0.912 & 0.887 & 0.925 & 0.940 & 0.936 & 0.934 & 0.938 \\ 
%\multicolumn{13}{c}{$r=0.5$, $b_{1}=0$, misspecification}\\
%  50 & 1.009 & 1.093 & 0.986 & 1.299 & 0.773 & 0.768 & 0.598 & 0.900 & 0.908 & 0.915 & 0.870 & 0.933 \\ 
%  100 & 1.371 & 1.502 & 1.379 & 1.549 & 0.805 & 0.954 & 0.779 & 0.924 & 0.924 & 0.946 & 0.921 & 0.942 \\ 
%  200 & 1.526 & 1.567 & 1.516 & 1.592 & 0.897 & 0.965 & 0.888 & 0.925 & 0.938 & 0.953 & 0.936 & 0.944 \\ \smallskip
%  500 & 1.576 & 1.587 & 1.569 & 1.595 & 0.913 & 0.937 & 0.911 & 0.912 & 0.943 & 0.949 & 0.944 & 0.941 \\ 
\multicolumn{13}{c}{$r=0.7$, $b_{1}=0$, misspecification}\\
  50 & 0.896 & 0.009 & 0.009 & 1.468 & 0.843 & 0.005 & 0.009 & 0.857 & 0.904 & 0.777 & 0.808 & 0.906 \\ 
  100 & 1.096 & 1.258 & 1.152 & 1.533 & 0.724 & 0.754 & 0.637 & 0.837 & 0.911 & 0.903 & 0.878 & 0.917 \\ 
  200 & 1.390 & 1.457 & 1.398 & 1.570 & 0.861 & 0.894 & 0.816 & 0.898 & 0.929 & 0.938 & 0.920 & 0.933 \\ \smallskip
  500 & 1.591 & 1.632 & 1.612 & 1.648 & 0.980 & 1.003 & 0.976 & 0.981 & 0.948 & 0.949 & 0.948 & 0.949 \\ 
\hline
\end{tabular}
}
\end{table}

\subsection{Simulation Studies with Binary Outcomes}

We also perform a set of simulations with binary outcomes, generating from a logistic outcome model. Three estimands, $\tau_{\RD},\tau_{\RR}$ and $\tau_{\COR}$, are considered in scenarios with different degree of treatment effect heterogeneity, prevalence of the outcome $\Pr(Y_{i}=1)$, and randomization probability $r$. In these scenarios, we find that covariate adjustment improves efficiency of the unadjusted estimator most likely when the sample size is at least $100$, except under large treatment effect heterogeneity where there is efficiency gain even with $N=50$. Throughout, the OW estimator is uniformly more efficient than IPW and should be the preferred propensity score weighting estimator in randomized trials. Finally, although the correctly-specified outcome regression is slightly more efficient than OW in the ideal case with a non-rare outcome, in small samples regression adjustment is generally unstable when the prevalence of outcome also decreases. Similarly, the efficiency of AIPW is mainly driven by the outcome regression component, and the instability of the outcome model may also lead to an inefficient AIPW estimator in finite-samples. For brevity, we present full details of the simulation design in Web Appendix D, and summarize all numerical results in Web Table 2 and 3. 

\section{Application to the Best Apnea Interventions for Research Trial} \label{sec:application}
The Best Apnea Interventions for Research (BestAIR) trial is an individually-randomized, parallel-group trial designed to evaluate the effect of continuous positive airway pressure (CPAP) treatment on the health outcomes of patients with high cardiovascular disease risk and obstructive sleep apnea but without severe sleepiness.\cite{Bakker2016} Patients were recruited from outpaient clinics at three medical centers in Boston, Massachusetts, and were randomized in a 1:1:1:1 ratio to receive conservative medical therapy (CMT), CMT plus sham CPAP (sham CPAP is a modified device that closely mimics the active CPAP and serves as the placebo for CPAP RCTs\cite{reid2019role}), CMT plus CPAP, or CMT plus CPAP plus motivational enhancement (ME). We follow the study protocol and pool two sub-arms into the combined control group (CMT, CMT plus sham CPAP) and the rest sub-arms into the combined CPAP or active intervention group. This results in 169 participants with 83 patients in the active CPAP group and 86 patients in the combined control arm. A set of patient-level covariates were measured at baseline and outcomes were measured at baseline, 6, and 12 months. 

For illustration, we consider estimating the treatment effect of CPAP on two outcomes measured at 6 month. The objective outcome is the 24-hour systolic blood pressure (SBP), measured every 20 minutes during the daytime and every 30 minutes during the sleep. The subjective outcome includes the self-reported sleepiness in daytime, measured by Epworth Sleepiness Scale (ESS). \cite{zhao2017effect} We additionally consider dichotomizing SBP (high SBP if $\geq$130mmHg) to create a binary outcome, resistant hypertension. For covariate-adjusted analysis, we consider a total of 9 baseline covariates, including demographics (e.g. age, gender, ethnicity), body mass index, Apnea-Hypopnea Index (AHI), average seated radial pulse rate (SDP), site and baseline outcome measures (e.g. baseline blood pressure and ESS). 

In Table \ref{tab:application_balance_check}, we provide the summary statistics for the covariates and compare between the treated and control groups at baseline. We measure the baseline imbalance of the covariates by the absolute standardized difference (ASD), which for the $j$th covariate is defined as, $\textup{ASD}^w=|\sum_{i=1}^{N}w_{i}X_{ij}Z_{i}/\sum_{i=1}^{N}w_{i}Z_{i}-\sum_{i=1}^{N}w_{i}X_{ij}(1-Z_{i})/\sum_{i=1}^{N}w_{i}(1-Z_{i})|/S_{j}$, where $w_{i}$ represents the weight for each patient and $S_{j}^2$ stands for the average variance, $S_{j}^{2}=\{\textup{Var}(X_{ij}|Z_{i}=1)+\textup{Var}(X_{ij}|Z_{i}=0)\}/2$. The baseline imbalance is measured by $\textup{ASD}^{\DIF}$ with $w_{i}=1$. Although the treatment is randomized, we still notice a considerable difference for some covariates between the treated and control group, such as BMI, baseline SBP and AHI. The $\textup{ASD}^{\DIF}$ for all three variables exceed $10\%$, which has been considered as a common threshold for balance.\cite{Austin2015} In particular, the baseline SBP exhibits the largest imbalance ($\textup{ASD}^{\DIF}=0.477$), and is expected to be highly correlated with SBP measured at 6 months, the main outcome of interest. As we shall see later, failing to adjust for such a covariate leads to spurious conclusions of the treatment effect. Using the propensity scores estimated from a main-effects logistic model, IPW reduces the baseline imbalance as $\textup{ASD}^{\IPW}<10\%$. As expected from the exact balance property (equation \eqref{eq:balance}), OW completely remove baseline imbalance such that $\textup{ASD}^{\OW}=0$ for all covariates. In this regard, even before observing the 6-month outcome, applying OW mitigates the severe imbalance on prognostic baseline factors, and thus increases the face validity of the trial. 

% Especially for the SBP measured at baseline period, which is strongly related to outcome measured after 6 months, the  $\textup{ASD}_{\DIF}$ is about $0.47$, with the control group has a higher SBP before the trial. As we shall see later, ignoring such imbalance in the covariates can lead to a biased result. we can plug in the weights from the IPW or OW method to calculate the ASD after weighting adjustment, such as $\textup{ASD}_{\IPW}$ and $\textup{ASD}_{\OW}$. Table \ref{tab:application_balance_check} shows that both IPW and OW greatly improve the balance of all covariates, while OW, as expected due to the exact balance property \eqref{eq:balance}, reduces the $\textup{ASD}_{\OW}$ to exactly zero for every covariate.  

\begin{table}[htbp]
    \centering
    \caption{\label{tab:application_balance_check} Baseline characteristics of the BestAIR randomized trial by treatment groups, and absolute standardized difference (ASD) between the treatment and control groups before and after weighting. The $\textup{ASD}^{\OW}$ is exactly zero due to the exact balance property of OW.}
    \begin{tabular}{lrrrrrr}
 \hline
 & All patients& CPAP group & Control group & $\textup{ASD}^{\DIF}$ & $\textup{ASD}^{\IPW}$ &$\textup{ASD}^{\OW}$ \\ 
 &$N=169$&$N_1=83$&$N_0=86$&&&\\
  \hline
 \multicolumn{7}{l}{Baseline categorical covariates and number of units in each group.}\\
  \smallskip
  Gender (male) & 107 & 54  & 53 & 0.046 & 0.002 & 0.000 \\ 
   \multicolumn{7}{l}{ Race \& ethnicity}\\
  \ \ White & 152 & 75  & 77  & 0.051 & 0.015 & 0.000 \\ 
  \ \ Black&  11 & 5  & 6  & 0.060 & 0.007 & 0.000 \\ \smallskip
  \ \ Other&  5 & 2 & 3  &  0.086 & 0.034 & 0.000\\
   \multicolumn{7}{l}{ Recruiting center}\\
  \ \ Site 1 & 54  & 26  & 28  & 0.046 & 0.002 & 0.000 \\ 
  \ \ Site 2 & 10  & 5   & 5 & 0.065 & 0.024 & 0.000 \\\smallskip
  \ \ Site 3 & 105 & 52  & 53  &0.073 & 0.013 & 0.000\\
  \hline
 \multicolumn{7}{l}{Baseline continuous covarites, mean and standard deviation (in parenthesis).}\\
  Age (years) & 64.4 (7.4)\  & 64.4 (8.0)\  & 64.3 (6.8) & 0.020 & 0.017 & 0.000 \\ 
 BMI (kg/$\textup{m}^2$) & 31.7 (6.0)\  & 31.0 (5.3)\  & 32.4 (6.5) & 0.261 & 0.042 & 0.000 \\ 
 Baseline SBP (mmHg) & 124.3  (13.2) & 121.6  (11.1) & 127.0 (14.6) & 0.477 & 0.020 & 0.000 \\ 
 Baseline SDP (beats/minute) & 63.1 (10.7) & 63.0 (10.4) & 63.2 (10.9) & 0.020 & 0.016 & 0.000 \\ 
 Baseline AHI (events/hr) & 28.8 (15.4) & 26.5 (13.0) & 31.1 (17.2) & 0.348 & 0.039 & 0.000 \\ 
 Baseline ESS & 8.3 (4.5) & 8.0 (4.5) & 8.5 (4.6) & 0.092 & 0.010 & 0.000 \\ 
   \hline
    \end{tabular}
\end{table}

For the continuous outcomes (SBP and ESS), we estimate the ATE using $\hat{\tau}^{\DIF}$, $\hat{\tau}^{\IPW}$, $\hat{\tau}^{\AIPW}$, $\hat{\tau}^{\LR}$ and $\hat{\tau}^{\OW}$. For IPW and OW, we estimate the propensity scores using a logistic regression with main effects of 9 baseline covariates mentioned above. For $\hat{\tau}^{\LR}$, we fit the ANCOVA model with main effects of treatment and covariates as well as their interactions. For the binary SBP, we use these five approaches to estimate the causal risk difference, log risk ratio and log odds ratio due to the CPAP treatment. For $\hat{\tau}^{\LR}$ with a binary outcome, we fit a logistic regression model for the outcome including both main effects of the treatment and covariates, as well as their interactions, and then obtain the marginal mean of each group via standardization. For each outcome, the corresponding propensity score and outcome model specifications are used to obtain the AIPW estimator. The variances and 95\% CIs of the estimators are calculated in the same way as in the simulations. We present p-values for the associated hypothesis tests of no treatment effect and occasionally interpret statistical significance at the 0.05 level for illustrative purposes. We do acknowledge, however, that the interpretation of study results should not rely on a single dichotomy of a p-value that is great than or smaller than 0.05.

\begin{table}[htbp]
\centering
\caption{\label{tab:application_result} Treatment effect estimates of CPAP intervention on blood pressure, day time sleepiness and resistant hypertension using data from the BestAIR study. The five approaches considered are: (a) UNADJ: the unadjusted estimator; %for binary outcome with a ratio estimand, this corresponds to the unadjusted difference-in-means estimator; 
(b) IPW: inverse probability weighting; (c) LR: linear regression (for continous outcomes, or ANCOVA) and logistic regression (for binary outcomes) for outcome;  (d) AIPW: augmented IPW; (e) OW: overlap weighting.}
\begin{tabular}{lcccc}
  \hline
Method  & Estimate & Standard error & 95\% Confidence interval & p-value \\ 
\hline
\multicolumn{5}{c}{Continuous outcomes}\\
  \multicolumn{5}{c}{Systolic blood pressure (continuous)}\\
UNADJ & $-5.070$ & $2.345$ & $(-9.667,-0.473)$ & $0.031$ \\ 
  IPW & $-2.638$ & $1.634$ & $(-5.841,\ \ 0.566)$ & $0.107$ \\ 
  LR & $-2.790$ & $1.724$ & $(-6.169,\ \ 0.588)$ & $0.106$ \\ 
  AIPW & $-2.839$ & $1.642$ & $(-6.058,\ \ 0.380)$ &$0.084$ \\ 
  OW & $-2.777$ & $1.689$ & $(-6.088,\ \  0.534)$ & $0.100$ \\ 
  \multicolumn{5}{c}{Epworth Sleepiness Scale (continuous)}\\
  UNADJ & $-1.503$ & $0.702$ & $(-2.878,-0.128)$ & $0.032$ \\ 
  IPW & $-1.232$ & $0.486$ & $(-2.184,-0.279)$ & $0.011$ \\ 
  LR & $-1.260$ & $0.519$ & $(-2.276,-0.243)$ & $0.015$ \\ 
  AIPW & $-1.255$ & $0.479$ & $(-2.193, -0.317)$ & $0.009$ \\ 
  OW & $-1.251$ & $0.491$ & $(-2.214,-0.288)$ & $0.011$ \\ 
 \multicolumn{5}{c}{Binary outcomes}\\
  \multicolumn{5}{c}{Resistant hypertension (SBP$\geq$130): risk difference}\\
  UNADJ & $-0.224$ & $0.085$ & $(-0.391,-0.057)$ & $0.009$ \\ 
  IPW & $-0.145$ & $0.082$ & $(-0.306,\ \  0.015)$ & $0.077$ \\ 
  LR & $-0.131$ & $0.074$ & $(-0.277,\ \  0.014)$ & $0.076$ \\ 
  AIPW & $-0.133$ & $0.071$ & $(-0.272,\ \  0.006)$ & $0.061$ \\ 
  OW & $-0.149$ & $0.083$ & $(-0.312,\ \ 0.013)$ & $0.071$ \\ 
  \multicolumn{5}{c}{Resistant hypertension (SBP$\geq$130): log risk ratio}\\
  UNADJ & $-0.698$ & $0.281$ & $(-1.248,-0.147)$ & $0.013$ \\ 
  IPW & $-0.448$ & $0.226$ & $(-0.892,-0.004)$ & $0.048$ \\ 
  LR & $-0.401$ & $0.236$ & $(-0.864,\ \ 0.062)$ & $0.090$ \\ 
  AIPW & $-0.408$ & $0.227$ & $(-0.854,\ \ 0.037)$ & $0.072$ \\ 
  OW & $-0.454$ & $0.263$ & $(-0.970,\ \ 0.062)$ & $0.084$ \\ 
  \multicolumn{5}{c}{Resistant hypertension (SBP$\geq$130): log odds ratio}\\
  UNADJ & $-1.038$ & $0.409$ & $(-1.838, -0.237)$ & $0.011$ \\ 
  IPW & $-0.665$ & $0.324$ & $(-1.300, -0.030)$ & $0.040$ \\ 
  LR & $-0.598$ & $0.346$ & $(-1.276,\ \ 0.080)$ & $0.084$ \\ 
  AIPW & $-0.607$ & $0.331$ & $(-1.256,\ \ 0.041)$ & $0.067$ \\ 
  OW & $-0.680$ & $0.387$ & $(-1.438,\ \ 0.079)$ & $0.079$ \\ 
   \hline
\end{tabular}
\end{table}

Table \ref{tab:application_result} presents the treatment effect estimates, standard errors (SEs), 95\% confidence intervals (CI) and p-values for these five approaches across three outcomes. For the SBP continuous outcome, the treatment effect estimated by IPW, LR, AIPW and OW are substantially smaller than the unadjusted estimate. Specially, the ATE changes from approximately $-5.0$ to $-2.7$ after covariate adjustment. This difference is due to the fact that the control group has a higher average SBP at baseline and failing to adjust for this discrepancy leads to a biased estimate of the treatment effect of CPAP. In fact, one would falsely conclude a statistically significant treatment effect at the $0.05$ level if the baseline imbalance is ignored. The treatment effect becomes no longer statistically significant at the $0.05$ level using either one of the adjusted estimator. In terms of efficiency, IPW, LR, AIPW and OW provide a smaller SE compared with the unadjusted estimate and the difference among the adjusted estimators is negligible. For the ESS outcome, the treatment effect estimate changes from approximately $-1.5$ to $-1.25$ after covariate adjustment while the difference among IPW, LR, AIPW and OW remains small. Despite the change in the point estimates, the 95\% confidence intervals for all five estimators exclude the null. 

For the binary SBP outcome, the unadjusted method gives an estimate of $-0.224$ on risk difference scale, $-0.698$ on log risk ratio scale and $-1.038$ on log odds ratio scale. Due to baseline imbalance, the unadjusted confidence intervals for all three estimands exclude null. Similar to the analysis of the continuous SBP outcome, all four adjusted approaches move the point estimates closer to the null. This pattern further demonstrates that ignoring baseline imbalance may produce biased estimates. In terms of variance reduction, all four adjusted methods exhibit a decrease in the estimated standard error compared with the unadjusted one. Interestingly, although the 95\% confidence intervals for LR, AIPW and OW all include zero, the confidence intervals for IPW excludes zero for the two ratio estimands (but not for the additive estimand). This result, however, needs to be interpreted with caution. As noticed in the simulation studies (panel (b), (c) and (d) in Web Figure 1), variance estimators of IPW and AIPW tend to underestimate the actual uncertainty when the sample size is modest and the outcome is not common. In our application, the resistant hypertension has a prevalence of around 12\%, which is close to the most extreme scenario in our simulation. Because IPW likely underestimates the variability for ratio estimands, there could be a risk of inflated type I error. In contrast, the interval estimator of OW appears more robust in small samples.

\section{Discussion} \label{sec:conclusion}
We advocate to use the overlap propensity score weighting (OW) method for covariate adjustment in randomized clinical trials. %Both OW and the previously proposed IPW belong to the general class of balancing weights. 
Compared with the regression adjustment approach, the propensity score approach encourages pre-planned adjustments of baseline covariates, and promote objectivity and transparency in the design and analysis of randomized trials. We have demonstrated that the OW estimator is asymptotically equivalent to the IPW and ANCOVA estimators, which becomes semiparametric efficient when the true outcome surface is linear in the covariates. 

Through extensive simulation studies, we find the OW estimator is consistently more efficient than the IPW estimator in finite samples, particularly when the sample size is small (e.g. smaller than 150). This is largely due to the exact balance property that is unique to OW, which removes all chance imbalance in the baseline covariates adjusted for in a logistic propensity model. Our simulations also shed light on the performance of the regression adjustment method. With a continuous outcome, linear regression adjustment have similar efficiency to the OW and IPW estimators when the sample size is more than $150$. With a limited sample size, say $N\leq 150$, the linear regression estimator is occasionally slightly more efficient than OW when correctly specified, while the OW estimator is more efficient when the linear model is incorrectly specified. We find that when the sample size is smaller than $100$, linear regression adjustment could even be less efficient than the unadjusted estimators when (i) the randomization probability deviates from $0.5$, and/or (ii) the outcome model is incorrectly specified. In contrast, the OW estimator consistently leads to finite-sample efficiency gain over the unadjusted estimator in these scenarios. Although we generally believe the sample size is a major determining factor for efficiency comparison, our cutoff of $N$ at $100$ or $150$ is specific to our simulation setting, and may not be generalizable to other scenarios we haven't considered. The findings for binary outcomes are slightly different from those for the continuous outcomes, especially in small samples (Web Appendix D). In particular, OW generally performs similarly to the logistic regression estimator, and both approaches may lead to efficiency loss over the unadjusted estimator when the sample size is limited, e.g., $N<100$. However, the efficiency loss generally does not exceed $10\%$. Throughout, the IPW estimator is the least efficient and could lead to over $20\%$ efficiency loss compared to the unadjusted estimator in small samples. The findings for estimating the risk ratio and odds ratio are mostly concordant with those for estimating the risk difference. Of note, when the binary outcome is rare, regression adjustment frequently run into convergence issues and fails to provide an adjusted treatment effect, while the propensity score weighting estimators are not subject to such problems. Finally, because previous simulations \cite{Moore2009,Moore2011,Colantuoni2015} with binary outcomes have focused on trials with at least a sample size of $N=200$, our simulations complement those previous reports by providing recommendations when the sample size falls below 200.  

We also empirically evaluated the finite-sample performance of the AIPW estimator in randomized trials. The AIPW estimator is popular in observational studies due to its double robustness and local efficiency properties. %That is, the AIPW estimator is consistent when either the propensity model or the outcome model is correctly specified, and becomes asymptotically more efficient than IPW when both models are correct. 
In randomized trials, because the propensity score model is never misspecified, the finite-sample performance of AIPW is largely driven by the outcome model. %matching the observations made in our simulation studies. 
In particular, we find that AIPW can be less efficient than the unadjusted estimator under outcome model misspecification (Table \ref{tab:cont_simu_results}). The sensitivity of AIPW to the outcome model specification was noted previously. \cite{li2013propensity,LiLi2019} AIPW could be slightly more efficient than OW with a correct outcome model and under substantial treatment effect heterogeneity, but it does not retain the objectivity of the simple weighting estimator and is subject to excessive variance when the outcome model is incorrect or fails to converge. 
%t appears to exacerbate the shortcomings of both IPW and outcome regression in many situations in randomized trials. 

We further provide a consistent variance estimator for OW when estimating both additive and ratio estimands. Our simulation results confirm that the resulting OW interval estimator achieved close to nominal coverage for the additive estimand (ATE), except in a few challenging scenarios where the sample size is extremely small, e.g. $N=50$. For example, with a continuous outcome, the empirical coverage of the OW interval estimator and the IPW interval estimator \cite{williamson2014variance} are both around $90\%$ when the randomization is unbalanced and the propensity score model does not account for important covariate interaction terms. In this case, the Huber-White variance for linear regression has the worst performance and barely achieved $80\%$ coverage. This is in sharp contrast to the findings of Raad et al.,\cite{Raad2020} who have demonstrated superior coverage of the linear regression interval estimator over the IPW interval estimator. However, Raad et al. \cite{Raad2020} only considered the model-based variance (i.e. based on the information matrix) when the outcome regression is correctly specified. Assuming a correct model specification, it is expected that the model-based variance has more stable performance than the Huber-White variance in small samples, while the former may become biased under incorrect model specification when the randomization probability deviates from $0.5$ .\cite{wang2019analysis} For robustness and practical considerations, we therefore focused on studying the operating characteristics of the commonly recommended Huber-White variance.\cite{lin2013agnostic} On the other hand, the OW interval estimator maintains at worst over-coverage for estimating the risk ratios and odds ratios when $N=50$, while the IPW interval estimator exhibits under-coverage. When the outcome is rare, the logistic regression and AIPW interval estimators show severe under-coverage possibly due to constant non-convergence. Collectively, these results indicate the potential type I error inflation by using IPW, logistic regression and AIPW, and could favor the application of OW for covariate adjustment in trials with a limited sample size. 

OW is easy to implement in practice. For applied researchers who are familiar with IPW, the switch to OW only involves a one-line change of the programming code: changing the weights from the reciprocal of the estimated probability of being assigned to the observed arm (IPW) to the probability of being assigned to the opposite arm (OW). Though the variance estimation is more complex, we have provided reproducible R code with implementation details in Web Appendix E and our GitHub page: \emph{\url{https://github.com/zengshx777/OWRCT\_codes\_package}}. Alternatively, OW, IPW and AIPW estimators are implemented in our recent R package \textbf{PSweight},\cite{zhou2020psweight} available on CRAN at \emph{\url{https://CRAN.R-project.org/package=PSweight}}. 

There are a number of possible extensions of the proposed method. First, subgroup analysis is routinely conducted in randomized trials to examine whether the treatment effect depends on certain sets of patient characteristics.\cite{Wang2007} For the same reason of transparency, it would be natural to develop propensity score weighting estimators for subgroup-specific treatment effects.\cite{Dong2020} Because the sample size of each subgroup may be limited, it is of particular interest to study whether OW is also effective in improving the efficiency in this context. Second, multi-arm randomized trials are common and the interest usually lies in determining the pairwise average treatment effect.\cite{Juszczak2019} Although the basic principle of improving efficiency via covariate adjustment still applies, there is a lack of empirical evaluation as to which adjustment approach works better in finite samples. In particular, the performance of multi-group ANCOVA and propensity score weighting merits further study. In the context of observational studies, we have previously extended OW to multiple treatments,\cite{li2019propensity} which is potentially applicable to multi-arm randomized trials. Third, although we have examined the AIPW estimator that combines IPW and direct regression, it remains to be explored whether an alternative hybrid estimator combining OW and outcome regression can lead to further improvement.\cite{li2020stabilizing} Lastly, covariate adjustment is also relevant in cluster randomized controlled trials, where entire clusters of patients (such as hospitals or clinics) are randomized to intervention conditions.\cite{turner2017a} Due to a limited number of clusters available in such studies, design-based adjustment for baseline characteristics are often considered by covariate-constrained randomization, \cite{li2016evaluation,li2017evaluation} in which case regression-based adjustment in the analysis stage is necessary not only for maintaining the type I error rate but also for efficiency improvement.\cite{Stephens2012} It remains an open question whether OW could similarly improve the performance of IPW for addressing challenges in the analysis of cluster randomized trials.

\section*{Appendix}
Web appendix is available at our GitHub page: \url{https://github.com/zengshx777/OWRCT_codes_package}.

\section*{Acknowledgments}

The authors gratefully acknowledge grants R01 AI136947 from the National Institute of Allergy and Infectious Diseases, 1U34HL105277 and 1R24HL114473 from National Heart, Lung, and Blood Institute. This work is also supported by CTSA Grant Number UL1 TR000142 from the National Center for Advancing Translational Science (NCATS), a component of the National Institutes of Health (NIH). Its contents are solely the responsibility of the authors and do not necessarily represent the official view of NIH.

\section*{Data Availability Statement}
The BestAIR trial data used in Section \ref{sec:application} are available upon reasonable request at \url{https://sleepdata.org}. 

%\subsection*{Conflict of interest}
%The authors declare no potential conflict of interests.

%\section*{Supporting information}
%The following supporting information is available as part of the online article:
%\noindent
%\textbf{Table A1.}

%\nocite{*}% Show all bib entries - both cited and uncited; comment this line to view only cited bib entries;
%\bibliography{wileyNJD-AMA}%
\bibliography{OW-RCT}
\end{document}